\documentstyle[aps,feynmf,epsfig]{revtex}
\begin{document}

\draft
\author{ M.Kiselev}
\address{Institut f\"ur Theoretische Physik, Universit\"at W\"urzburg,
D-97074 W\"urzburg, Germany\\}
\date{\today}

\title{Semi-fermionic approach for quantum spin systems}
\maketitle
\vspace*{-3cm}
\hspace*{5cm}
{\it Extended version of the talk given at TH-2002 conference, Paris, July 22-27, 2002}\\
\vspace*{2cm}
\begin{abstract}
We present a general derivation of semi-fermionic representation for generators of $SU(N)$ group
as a bilinear combination of Fermi operators.
The constraints are fulfilled by means of imaginary Lagrange multipliers. The important case
of $SU(2)$ group is discussed. We demonstrate how the idea of semi-fermionic representation
might be extended to the groups possessing dynamic symmetries. 
As an example, $SO(4)$ group is considered.
We illustrate the application of semi-fermionic representations for various problems of strongly 
correlated physics.\\
\mbox{}\\
\mbox{}
\end{abstract}

\section*{Introduction}
It is known that spin operators satisfy neither Fermi nor Bose commutation relations. For example,
the Pauli matrices for $S=1/2$ operator commute on different sites and anticommute on the same site.
The commutation relations for spins are determined by $SU(2)$ algebra, leading to the absence of a
Wick theorem for the generators. To avoid this difficulty and construct a diagrammatic
technique and path integral representation for spin systems various approaches have been used.
The first class of approaches is based on representation of spins as bilinear combination of
Fermi or Bose operators \cite{holstein40a}-\cite{larkin68b}, whereas the representations 
belonging to the second class deal with
more complex objects like, e.g. the Hubbard \cite{hubbard65a} and supersymmetric 
\cite{coleman00a} operators, the 
nonlinear sigma model \cite{book2} etc. However, in all cases
the fundamental problem which is  at the heart of the difficulty is the local constraint problem.
To illustrate it, let's consider e.g., first class of representations. Introducing the 
auxiliary Fermi or Bose fields
makes the dimensionality of the Hilbert space, where these operators act, greater than the 
dimensionality of the Hilbert space for the spin operators. As a result, the spurious unphysical states 
should be excluded from the consideration which leads in turn  to some restrictions (constraints) 
on bilinear combinations of Fermi/Bose operators, resulting in substantial complication
of corresponding rules of the diagrammatic technique. 
The representations from the second class suffer from the 
same kind of problem, transformed either into a high nonlinearity of resulting model 
(non-linear sigma model) or hierarchical structure of perturbation series in the absence of Wick theorem
(Hubbard operators). The exclusion of double occupied and empty states for a $S=1/2$ impurity
interacting with conduction electron bath (single impurity Kondo model), is controlled by fictitious 
chemical potential (Lagrange multiplier) of Abrikosov pseudofermions \cite{abrikosov65a}. 
At the end of calculations
this ``chemical potential'' $\lambda$ should be put $\lambda \to -\infty$ to ``freeze out''
all unphysical states. In other words, there exists an additional $U(1)$ gauge field which freezes
the charge fluctuations associated with this representation. The method 
works for dilute systems where all the spins
can be considered independently. Unfortunately, attempts to generalize this technique to the lattice
of spins results in the replacement of the local constraint (the number of particles on each site is
fixed) by the so-called global constraint where the number of particles is fixed only on an average
for the whole crystal. There is no reason to believe that such an approximation is a good starting 
point for the description of the strongly correlated systems.
Another possibility to treat the local constraint rigorously is based on
Majorana fermion representation. In this case fermions are ``real'' and corresponding gauge symmetry is 
$Z_2$. The difficulty with this representation is mostly related to the physical regularization
of the fluctuations associated with the discrete symmetry group.

An alternative approach for spin Hamiltonians, free from local constraint problem,
has been proposed in the pioneering paper of Popov and Fedotov \cite{popov88a}. Based on the exact
fermionic representation for $S=1/2$ and $S=1$ operators, where the constraint is controlled by
purely imaginary Lagrange multipliers, these authors demonstrated the power and simplification
of the corresponding Matsubara diagram technique. The semi-fermionic representation
(we discuss the meaning of this definition in the course of our paper) used by Popov and Fedotov
is neither fermionic, nor bosonic, but reflects the fundamental Pauli nature of spins. 
The goal of this paper is to give a brief introduction  to a semi-fermionic (SF) approach. A reader can
find many useful technical details, discussion of mathematical aspects
of semi-fermionic representation and its application to various problems in the 
original papers \cite{popov88a}-\cite{col02}. 
However, we reproduce the key steps of important derivations 
contained in \cite{kis01},\cite{kis02a} in order to make the reader's job easier. 

The manuscript is organized as follows: in Section I, the general concept of semi-fermions
is introduced. We begin with the construction of the SF formalism for the 
fully antisymmetric representation 
of $SU(N)$ group
and the fully symmetric SF representation of $SU(2)$ group using the 
imaginary-time (Matsubara) representation. 
We show a ``bridge'' between different representations using the simplest example of $S=1$ in $SU(2)$
and discuss the SF approach for 
$SO(4)$ group. Finally, we show how to work with semi-fermions in real-time formalism and construct
the Schwinger-Keldysh technique for SF. In this section, we will mostly follow original papers 
by the author \cite{kiselev00b}, \cite{kis01}.
The reader acquainted with semi-fermionic technique can easily skip this section.
In Section II, we illustrate the applications of SF formalism for various problems of 
condensed matter physics, such  as ferromagnetic (FM), antiferromagnetic (AFM) 
and resonance valence bond (RVB) instabilities in the Heisenberg model, competition between
local and non-local correlations in Kondo lattices 
in the vicinity of magnetic and spin glass critical points and the Kondo effect in quantum dots. 
In the Epilogue, we discuss some open questions and perspectives.
\section{Semi-fermionic representation}
To begin with, we briefly reproduce the arguments contained in the 
original paper of Popov and Fedotov.
Let's assume first $S=1/2$. We denote as 
$H_\sigma$ the Hamiltonian of spin system. The standard Pauli 
matrices can be represented as bilinear combination of Fermi operators as follows:
\begin{equation}
\sigma_j^z\to a_j^\dagger a_j - b_j^\dagger b_j,\;\;\;\;\;\;
\sigma_j^+\to 2 a_j^\dagger b_j,\;\;\;\;\;\;\;
\sigma_j^-\to 2 b_j^\dagger a_j.
\label{rep1}
\end{equation}
on each site $i$ of the lattice. The partition function of the spin problem $Z_\sigma$
is given by
\begin{equation}
Z_{\sigma}= Tr \exp(-\beta \hat H_\sigma)=i^N Tr\exp(-\beta (\hat H_F+i\pi \hat N_F/(2\beta))
\label{pf1}
\end{equation}
where $\hat H_F$ is the operator obtained from $\hat H_\sigma$ by the replacement (\ref{rep1}) and
\begin{equation}
\hat N=\sum_{j=1}^N(a^\dagger_j a_j +b^\dagger_j b_j)
\end{equation}
($N$ is the number of sites in the system and $\beta=1/T$ is inverse temperature). To prove equation 
(\ref{pf1})
we note that the trace over the nonphysical states of the $i$-th site vanishes
\begin{equation}
Tr_{unphys}\exp(-\beta (\hat H_F+i\pi \hat N_F/(2\beta))= (-i)^0+(-i)^2=0
\end{equation}
Thus, the identity (\ref{pf1}) holds. The constraint of fixed number of fermions $\hat N_j=1$,
 is achieved by means of the
purely imaginary Lagrange multipliers $\mu=-i\pi/(2\beta)$ playing the role of imaginary chemical 
potentials of fermions. As a result, the Green's function
\begin{equation}
G=(i\omega_F-\epsilon)^{-1}
\end{equation}
is expressed in terms of  Matsubara frequencies $\omega_F=2\pi T(n+1/4)$  corresponding 
neither Fermi nor Bose statistics.

For $S=1$ we adopt the representation of $\hat H_\sigma$ in terms of the 3-component Fermi field:
\begin{equation}
\sigma^z_j\to a^\dagger_j a -b^\dagger_j b,\;\;\;\;\;
\sigma^+_j\to \sqrt{2}(a_j^\dagger c_j +c_j^\dagger b_j),\;\;\;\;
\sigma^-_j\to \sqrt{2}(c_j^\dagger a_j +b_j^\dagger c_j).
\label{s1}
\end{equation}
The partition function $Z_\sigma$ is given by
\begin{equation}
Z_{\sigma}=Tr (-\beta \hat H_\sigma)=\left(\frac{i}{\sqrt{3}}\right)^N 
Tr\exp(-\beta (\hat H_F+i\pi \hat N_F/(3\beta)).
\end{equation}
It is easy to note that the states with occupation numbers 0 and 3 cancel each other, whereas
states with occupation 1 and 2 are equivalent due to the particle-hole symmetry and thus can be taken
into account on an equal footing by proper normalization of the partition function. As a result,
the Green's function in the 
imaginary time representation is expressed in terms of $\omega_F=2\pi T(n+1/3)$
frequencies.

In this section, we show how semi-fermionic (Popov-Fedotov) representation can be derived
using the mapping of partition function of the spin problem onto the corresponding partition function of 
the fermionic problem. The cases of arbitrary N (even) for SU(N) groups and arbitrary S for SU(2)
group are discussed.
\subsection{SU(N) group}
We begin with the derivation of SF representation for SU(N) group.
The SU($N$) algebra is determined by the generators obeying the
following commutation relations:
\begin{equation}
[\hat S^\beta_{\alpha, i} \hat S^\rho_{\sigma j}]=
\delta_{ij}(\delta^\rho_\alpha \hat S^\beta_{\sigma i}-\delta^\beta_\sigma
\hat S^\rho_{\alpha i}),
\end{equation}
where $\alpha,\beta=1,...,N$. We adopt the definition of the
Cartan algebra \cite{cartan} of the SU($N$) group
$\{H_\alpha\}=S_\alpha^\alpha$ similar to the one used in \cite{sachdev89a},
noting that the diagonal generators $S_\alpha^\alpha$ are not traceless.
To ensure a vanishing trace, the
diagonal generators should only appear in combinations
\begin{equation}
\sum_{\alpha=1}^{N} s_\alpha S_\alpha^\alpha \quad \mbox{with} \quad
\sum_{\alpha=1}^{N} s_\alpha = 0,
\end{equation}
which effectively reduce the number of independent diagonal generators to
$N-1$ and the total number of SU($N$) generators to $N^2-1$.

In this paper we discuss the representations of SU(N) group determined by rectangular 
Young Tableau (YT)
(see \cite{sachdev89a} and \cite{kis01} for details) 
and mostly concentrate on two important cases of the fully asymmetric
(one column) YT and the fully symmetric (one row) YT.

The generator  $\hat S^\alpha_\beta$  may be written as biquadratic form
in terms of the Fermi-operators
\begin{equation}
\hat S^\alpha_\beta=\sum_\gamma a^\dagger_{\alpha \gamma}a^{\beta \gamma}
\label{fermi}
\end{equation}
where the "color" index $\gamma=1,...,n_c$ and the $n_c(n_c+1)/2$
constraints
\begin{equation}
\sum_{\alpha=1}^Na^\dagger_{\alpha \gamma_1} a^{\alpha \gamma_2}=\delta_{\gamma_{1}}^{\gamma_{2}} m
\label{gconst}
\end{equation}
restrict the Hilbert space to the states with $m * n_c$ particles and
ensure the characteristic symmetry in the color index $a$.  Here $m$ corresponds
to the number of rows in rectangular Young Tableau whereas $n_c$ stands for the number of columns.
The antisymmetric behavior with respect to $\alpha$ is a direct
consequence of the fermionic representation.

Let us consider the partition function for the Hamiltonian, expressed
in terms of SU($N$) generators
\begin{equation}
Z_{S}=Tr \exp(-\beta H_S)= Tr' \exp(-\beta H_F)
\label{z1}
\end{equation}
where $Tr'$ denotes the trace taken with constraints (\ref{gconst}).
As it is shown in \cite{kis01}, the partition function of $SU(N)$ model
is related to partition function of corresponding fermion model through
the following equation:
\begin{equation}
Z_S=\int\prod_{j} d\mu(j) P(\mu(j))
Tr\exp\left(-\beta(H_F-\mu(j)n_F)\right)
=\int\prod_{j} d\mu(j) P(\mu(j))Z_F(\mu(j))
\end{equation}
here $P(\mu_j)$ is a distribution function of imaginary Lagrange multipliers. We calculate
$P(\mu_j)$ explicitely using constraints (\ref{gconst}).

We use the path integral representation of the partition function
\begin{equation}
Z_S/Z_S^0=\int\prod_{j} d\mu(j) P(\mu(j))\exp({\cal A})/
\int\prod_{j} d\mu(j) P(\mu(j))\exp({\cal A}_0)
\label{z2}
\end{equation}
where the actions ${\cal A}$ and ${\cal A}_0$ are determined by
\begin{equation}
{\cal A}={\cal A}_0 - \int_0^\beta d\tau H_F(\tau),\;\;\;
{\cal A}_0=\sum_{j}\sum_{k=1}^N\int_0^\beta d\tau\bar a_k(j,\tau)
(\partial_\tau+\mu(j))a_k(j,\tau)
\end{equation}
and the fermionic representation of SU($N$) generators (\ref{fermi})
is applied.

Let us first consider the case $n_c=1$. We denote the corresponding
distribution by $P_{N,m}(\mu(j))$, where $m$ is the number of particles
in the SU($N$) orbital, or in other words, $1\leq m < N$ labels the
different fundamental representations of SU($N$).
\begin{equation}
n_j=\sum_{k=1}^N \bar a_{k}(j) a_{k}(j)=m
\label{const1}
\end{equation}
To satisfy this requirement, the minimal set of chemical potentials
and the corresponding form of $P_{N,m}(\mu(j))$ are to be derived.

To derive the distribution function, we use the following identity for the
constraint (\ref{const1}) expressed in terms of Grassmann variables
\begin{equation}
\delta_{n_j, m}=\frac{1}{N}
\sin\left(\pi(n_j-m)\right)/
\sin\left(\frac{\pi(n_j-m)}{N}\right)
\label{d1}
\end{equation}
Substituting this identity into (\ref{z1}) and comparing with
(\ref{z2}) one gets
\begin{equation}
P_{N,m}(\mu(j))=\frac{1}{N}\sum_{k=1}^{N}
\exp\left(\frac{i\pi m}{N}(2k-1)\right)\delta(\mu(j)-\mu_k),
\label{eq:P_v1}
\end{equation}
where
\begin{equation}
\mu_k = - \frac{i \pi T}{N}(2k-1).
\label{eq:mu_k}
\end{equation}
Since the Hamiltonian is symmetric under the exchange of particles and
holes when the sign of the Lagrange multiplier is also changed simultaneously,
we can simplify (\ref{eq:P_v1}) to
\begin{equation}
P_{N,m}(\mu(j))=
\frac{2 i}{N}\sum_{k=1}^{\lfloor N/2 \rfloor}
\sin\left(\pi m\frac{2k-1}{N}\right)\delta(\mu(j)-\mu_k)
\label{dfu}
\end{equation}
where $\lfloor N/2 \rfloor$ denotes the integer part of $N/2$.  As shown below, 
this is the minimal representation of the
distribution function corresponding to the minimal set of the discrete
imaginary Lagrange multipliers. Another distributions function
different from (\ref{dfu}) can be constructed when the sum
is taken from $k=N/2+1$ to $N$. Nevertheless, this DF is
different from (\ref{dfu}) only by the sign of imaginary Lagrange multipliers
$\tilde \mu_k=\mu_k^*=-\mu_k$ and thus is supplementary to (\ref{dfu}).

Particularly interesting for even $N$ is the case when the SU($N$)
orbital is half--filled, $m=N/2$.  Then all Lagrange multipliers carry  equal weight
\begin{equation}
P_{N,N/2}(\mu(j))=\frac{2i}{N}\sum_{k=1}^{N/2}(-1)^{k+1}\delta\left(\mu(j)-\mu_k\right).
\label{nn2}
\end{equation}
Taking the limit $N\to\infty$ one may replace the summation in
expression (\ref{nn2}) in a suitable way by integration.  Note, that while
taking $N\to\infty$ and $m\to\infty$ limits, we nevertheless keep the ratio
$m/N=1/2$ fixed. Then, the
following limiting distribution function
can be obtained:
\begin{equation}
P_{N,N/2}(\mu(j)) \stackrel{N\to\infty}{\longrightarrow}
\frac{\beta}{2\pi i}
\exp\left(-\beta \mu(j) \frac{N}{2}\right)
\label{pg}
\end{equation}
resulting in the usual continuous representation of the local
constraint for the simplest case $n_c=1$
\begin{equation}
Z_S=Tr(\exp\left(-\beta H_F\right) \delta \left(n_j -\frac{N}{2})\right)
\end{equation}
We note the obvious similarity of the limiting DF (\ref{pg}) with the
{\it Gibbs canonical distribution} provided that
the Wick rotation from the imaginary axis of the Lagrange multipliers $\mu$
to the real axis of energies $E$ is performed and thus $\mu(j) N/2$ has a
meaning of energy.

Up to now, the representation we discussed was purely fermionic and
expressed in terms of usual Grassmann variables when the path integral
formalism is applied. The only difference from slave fermionic
approach is that imaginary Lagrange multipliers are introduced to
fulfill the constraint.  Nevertheless, by making the replacement
\begin{equation}
a_k(j,\tau)) \to a_k(j,\tau)
  \exp\left(\frac{i\pi\tau}{\beta} \frac{2k-1}{N}\right),\;\;\;
  \bar a_k(j,\tau) \to \bar a_k(j,\tau)
  \exp\left(-\frac{i\pi\tau}{\beta} \frac{2k-1}{N}\right)
\end{equation}
we arrive at the generalized Grassmann (semi-fermionic) boundary
conditions
\begin{equation}
 a_k(j,\beta) = a_k(j,0)\exp\left(i\pi \frac{2k-1}{N}\right),\;\;\;
          \bar a_k(j,\beta) = \bar a_k(j,0)\exp\left(-i\pi
            \frac{2k-1}{N}\right)
\label{bk}
\end{equation}
This leads to a temperature diagram technique for the Green's functions
\begin{equation}
{\cal G}^{\alpha\beta}(j,\tau)=-
\langle T_\tau a_\alpha(j,\tau) \bar a_\beta(j,0)\rangle
\label{itf}
\end{equation}
of semi-fermions with Matsubara frequencies different from both Fermi
and Bose representations (see Fig.\ref{fig:frequencies_spin}).

\begin{figure}
\begin{center}
  \epsfysize4cm \epsfbox{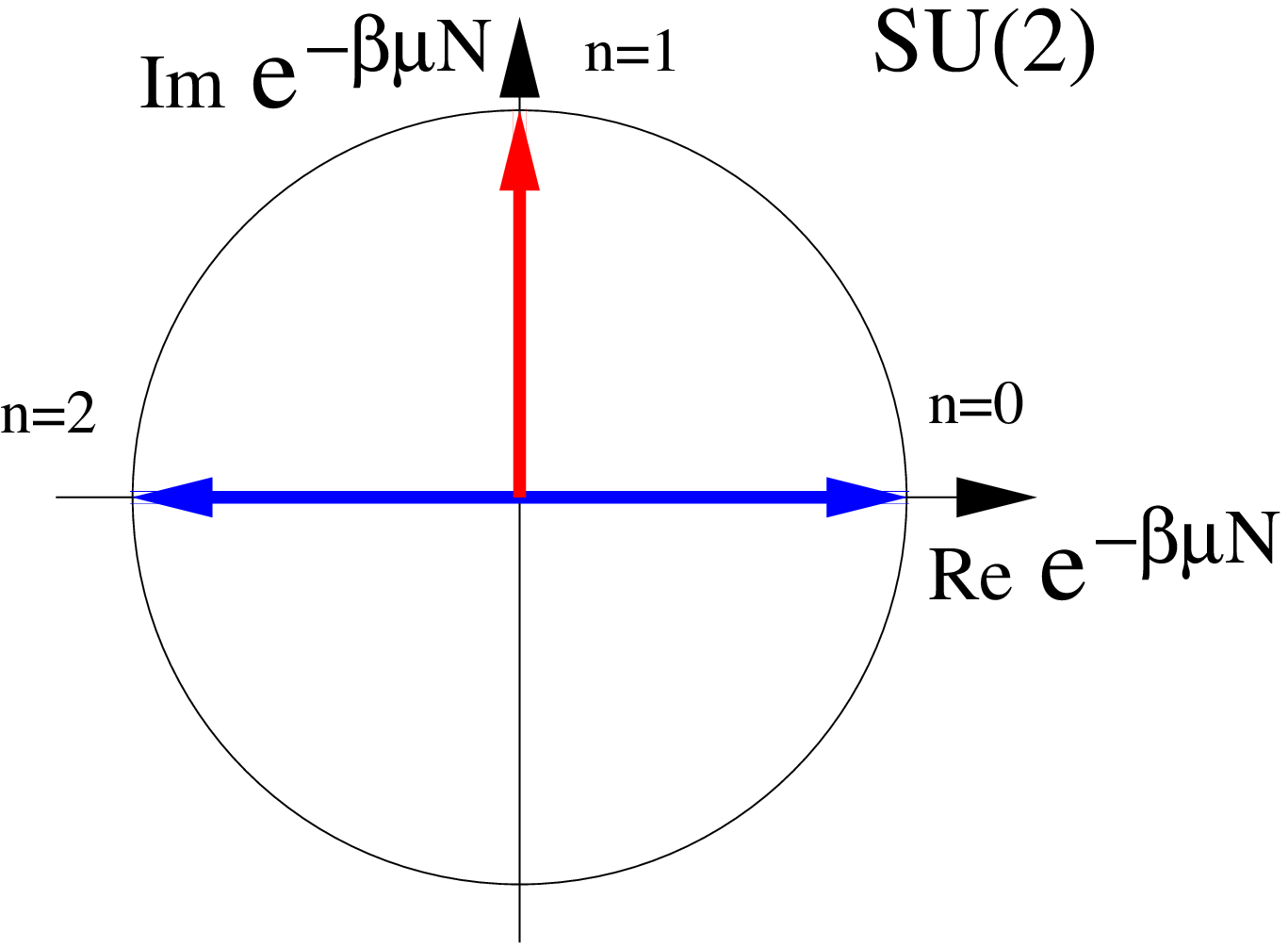}
  \epsfysize4cm
\epsfbox{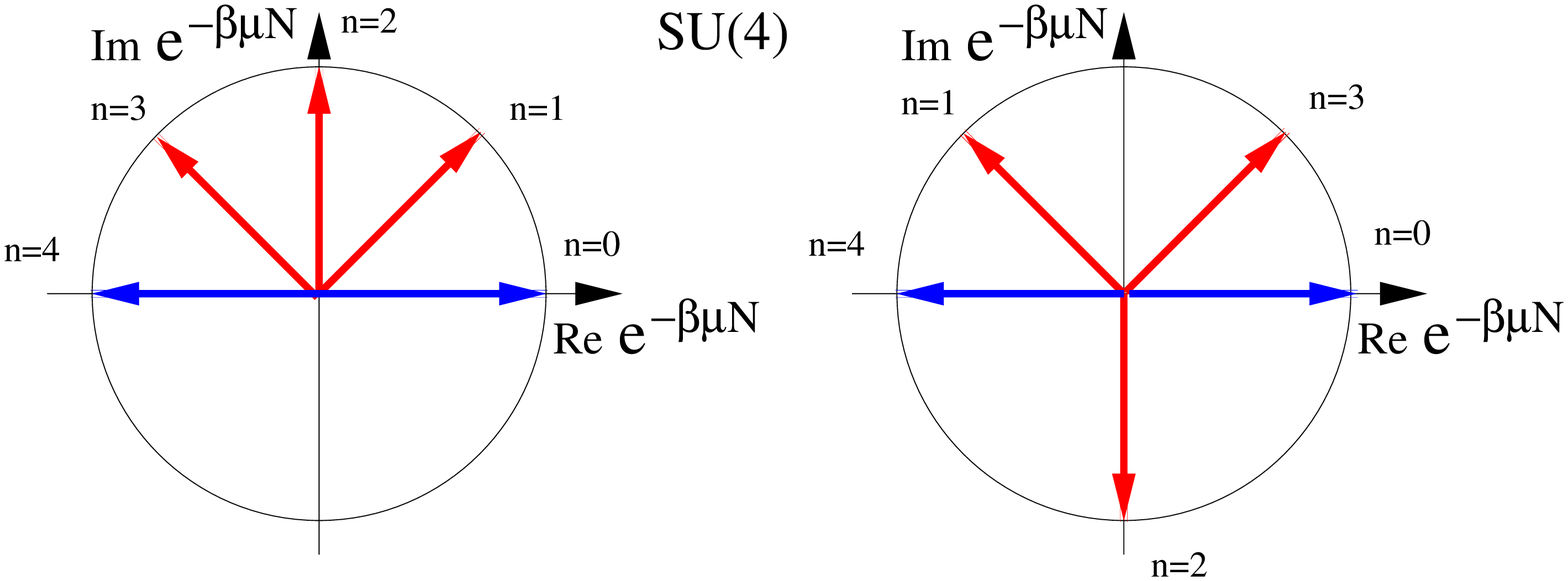}
\end{center}
\caption{Graphical representation of exclusion principle for SU($N$)
  semi-fermionic representation with even $N$, $n_c=1$
(we use $\mu=i\pi T/2$ for SU(2) and
$\mu_1=i\pi T/4,\;\mu_2=3 i \pi T/4$ for SU(4)).}
\label{fig:SU(N)_rep}
\end{figure}

The exclusion principle for this case is illustrated on
Fig.\ref{fig:SU(N)_rep}, where the $S=1/2$ representation for the first two groups SU(2) and SU(4)
are shown. The first point to observe is that the spin Hamiltonian
does not distinguish the $n$ particle and the $n$ hole (or $N-n$
particle) subspace. Eq. (\ref{eq:mu_k}) shows that the two phase factors
$\exp(\beta \mu n)$ and $\exp(\beta \mu (N-n))$ accompanying these
subspaces in Eq. (\ref{dfu}) add up to a purely imaginary value within the
same Lagrange multiplier, and the empty and the fully occupied states are
always canceled. In the case of $N \geq 4$, where we have multiple
Lagrange multipliers, the distribution function $P(\mu)$ linearly
combines these imaginary prefactors to select out the desired physical
subspace with particle number $n=m$.

In Fig.\ref{fig:SU(N)_rep}, we note that on each picture, the empty and
fully occupied states are canceled in their own unit circle. For SU(2)
there is a unique chemical potential $\mu=\pm i\pi T/2$ which results in
the survival of single occupied states. For SU(4) there are two chemical
potentials (see also Fig.\ref{fig:frequencies_spin}).
The cancellation of single and triple
occupied states is achieved with the help of proper weights for these
states in the distribution function whereas the states with the
occupation number 2 are doubled according to the expression
(\ref{nn2}). In general, for SU($N$) group with $n_c=1$ there exists
$N/2$ circles providing the realization of the exclusion principle.

\subsection{SU(2) group}

We consider now the generalization of the SU(2) algebra for the case of spin $S$.
Here, the most convenient fermionic representation is constructed with
the help of a $2S+1$ component Fermi field $a_k(j)$ provided that the
generators of SU(2) satisfy the following equations:
$$
 S^+=\sum_{k=-S}^{S-1}\sqrt{S(S+1)-k(k+1)}a^\dagger_{k+1}(j)a_k(j),\;\;
 S^-=\sum_{k=-S+1}^{S}\sqrt{S(S+1)-k(k-1)}a^\dagger_{k-1}(j)a_k(j),
$$
\begin{equation}
S^z=\sum_{k=-S}^S k a^\dagger_{k}(j)a_k(j)
\label{su2}
\end{equation}
such that $dim H_F=2^{2S+1}$ whereas the constraint reads as follows
\begin{equation}
n_j=\sum_{k=-S}^{k=S}a^\dagger_{k}(j) a_{k}(j)=l=1
\label{const2}
\end{equation}
 Following the same routine as for
SU($N$) generators and using the occupancy condition to have $l=1$ (or
$2S$) states of the $(2S+1)$ states filled, one gets the following
distribution function, after using the particle--hole symmetry of the Hamiltonian
$H_S$:
\begin{equation}
P_{2S+1,1}(\mu(j))=\frac{2i}{2S+1}\sum_{k=1}^{\lfloor S+1/2\rfloor}
\sin\left(\pi \frac{2k-1}{2S+1}\right)\delta(\mu(j)-\mu_k)
\label{lsu}
\end{equation}
where the Lagrange multipliers are $\mu_k=-i\pi T(2k-1)/(2S+1)$ and
$k=1,...,\lfloor S+1/2 \rfloor$, similarly to Eq.(\ref{eq:mu_k}).
\begin{figure}
\begin{center}
  \epsfxsize8cm \epsfysize5cm \epsfbox{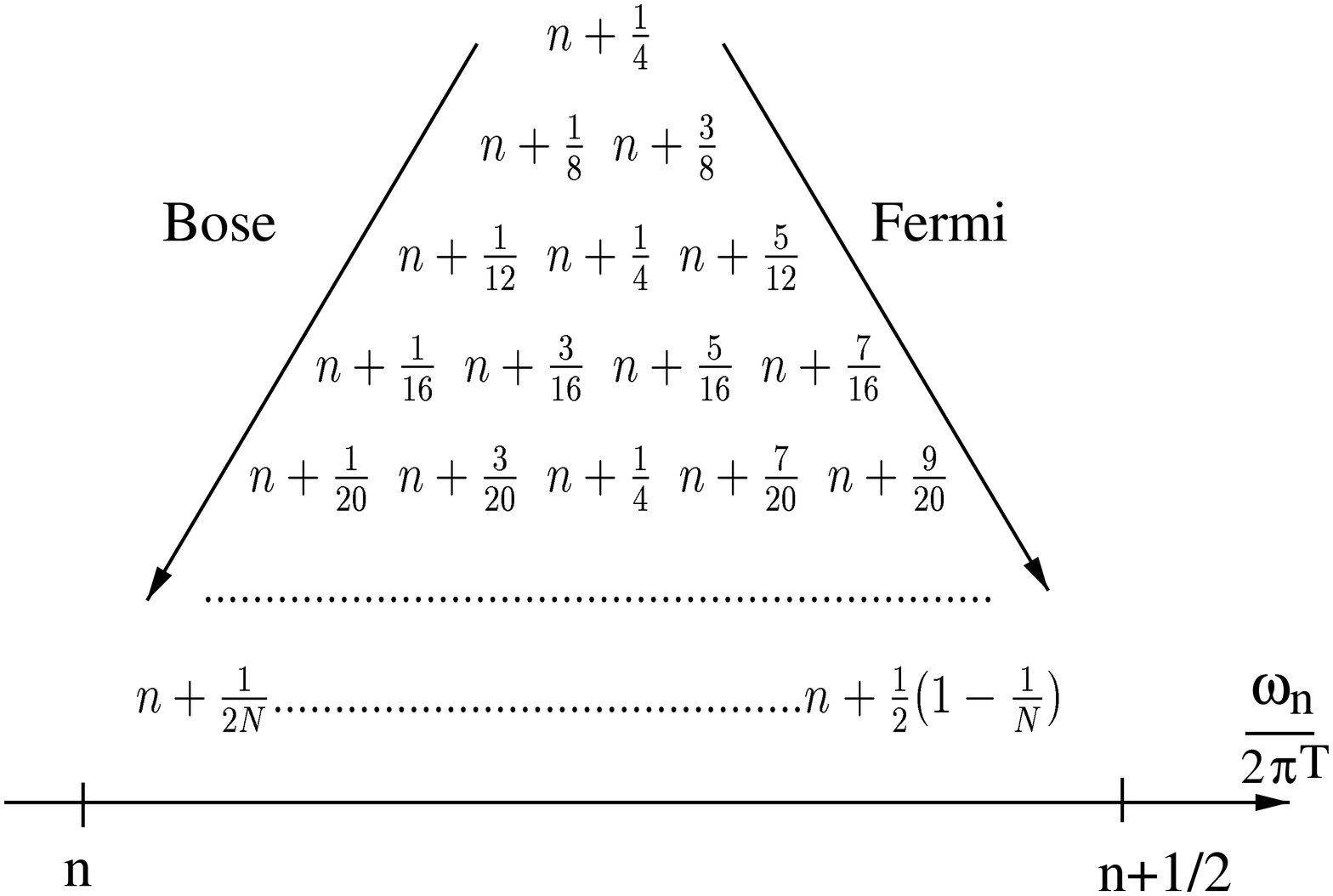}\hspace*{1cm}
  \epsfxsize8cm \epsfysize5cm \epsfbox{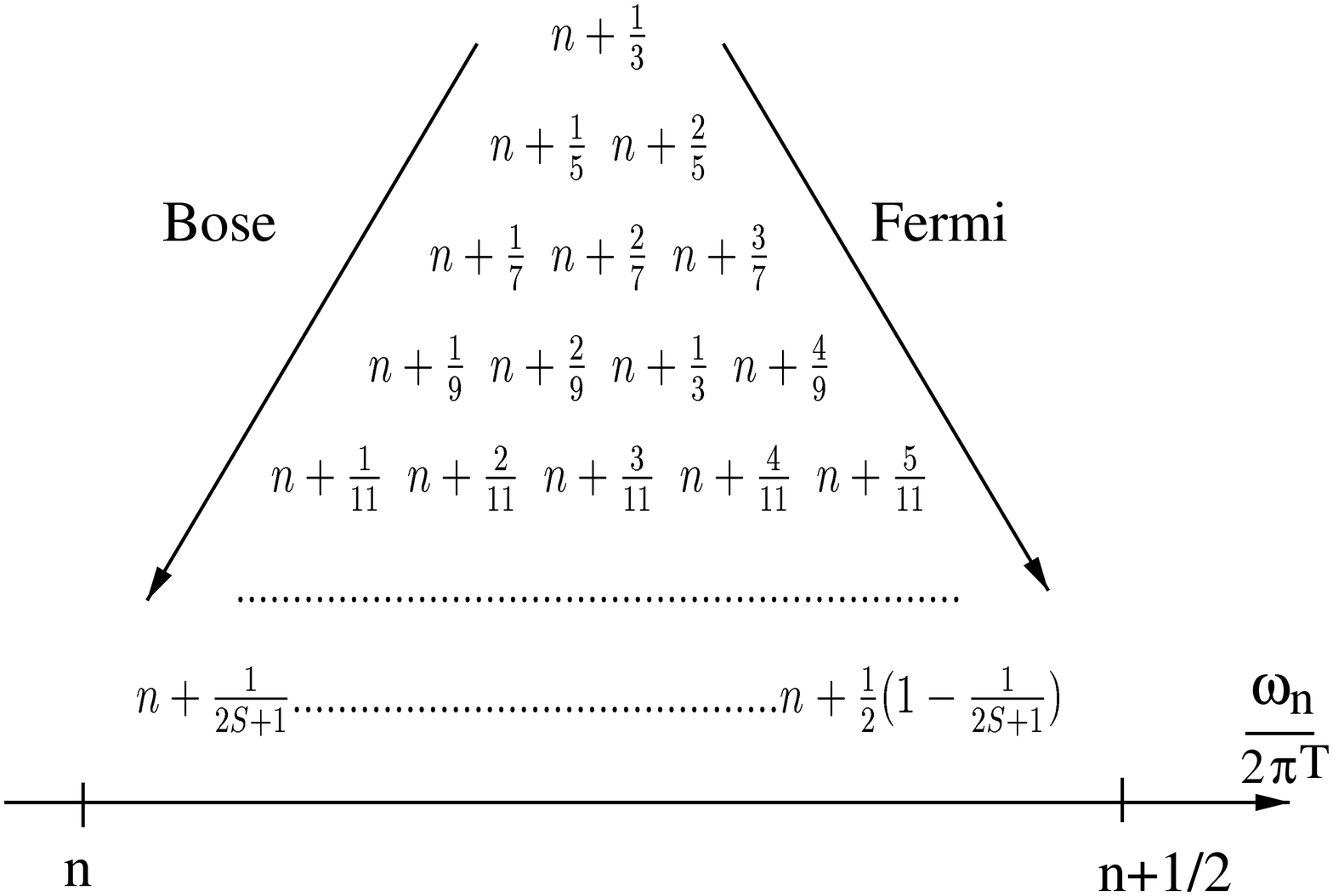}
\end{center}
\caption{The minimal set of Matsubara frequencies for 
a) $SU(N)$ representation with even $N$/ $SU(2)$ representation for half-integer value of the spin.  
b) $SU(2)$  representation for integer values of the spin and $l=1$.}
\label{fig:frequencies_spin}
\end{figure}
In the particular case of the SU(2) model  for some chosen
values of spin $S$ the distribution functions are given by the
following expressions
\begin{equation}
  P_{2,1}(\mu(j))=i\;\delta\left(\mu(j)+\frac{i\pi T}{2}\right)
\nonumber
\end{equation}
for $S=1/2$
\begin{equation}
 P_{3,1}(\mu(j))=P_{3,2}(\mu(j))=
  \frac{i}{\sqrt{3}}\;\delta\left(\mu(j)+\frac{i\pi T}{3}\right)
\nonumber
\end{equation}
for $S=1$.

This result corresponds to the original Popov-Fedotov description
restricted to the $S=1/2$ and $S=1$ cases.
\begin{figure}
\begin{center}
  \epsfysize4cm  \epsfbox{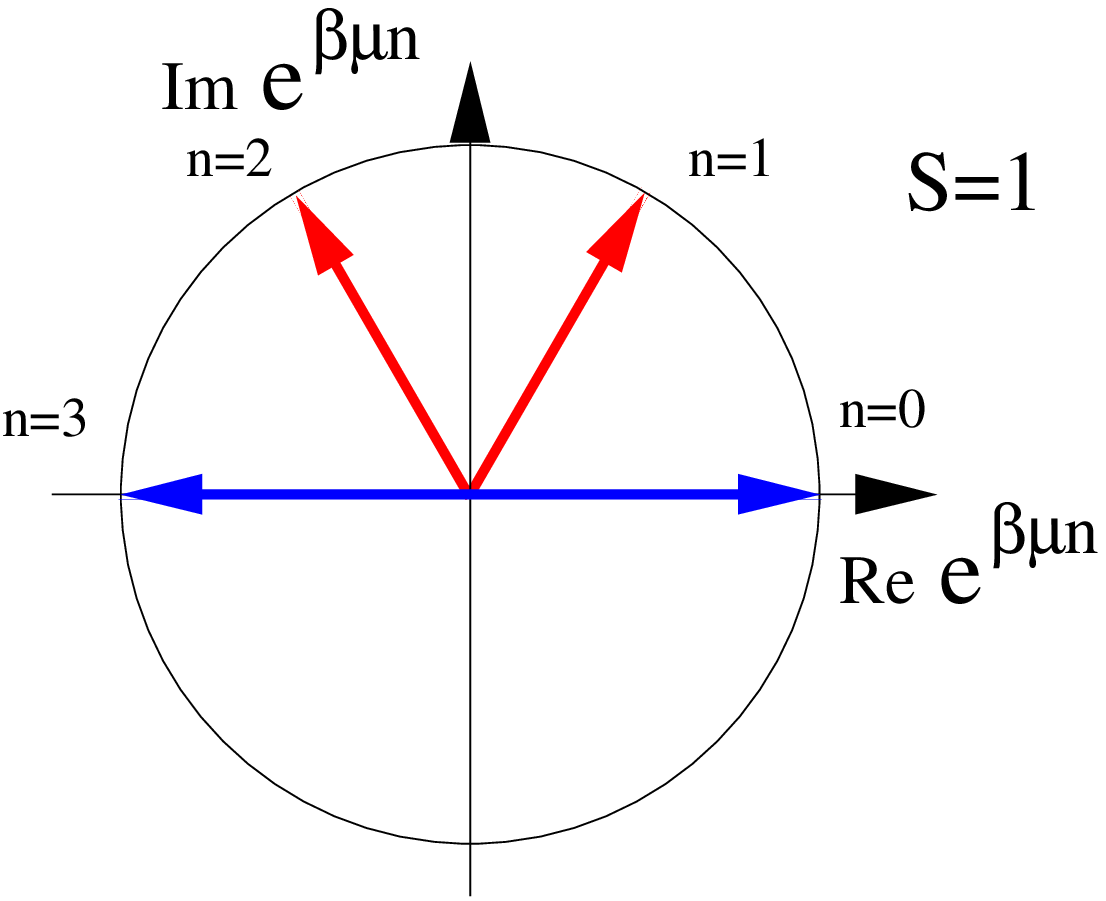}
   \epsfysize4cm \epsfbox{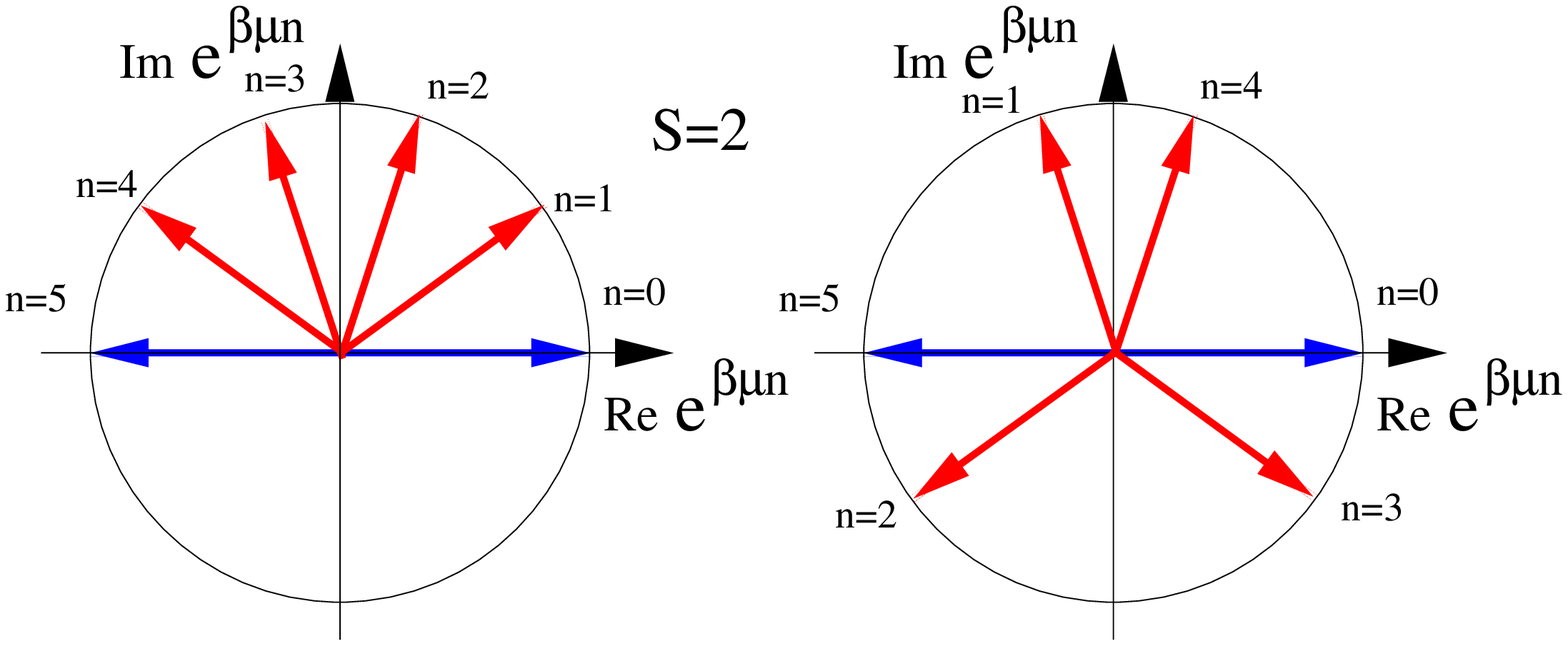}
\end{center}
\caption{Graphical representation of exclusion principle for SU(2)
  semi-fermionic representation for $S=1$  and $S=2$.
  For any arbitrary integer
  value of spin there exists $S$ circle diagrams corresponding to the
  $S$ different chemical potentials and providing the realization of
  the exclusion principle.}
\label{fig:spin_rep}
\end{figure}
A limiting distribution function corresponding to Eq. (\ref{pg}) for
the constraint condition with arbitrary $l$ is found to be
\begin{equation}
P_{\infty,l}(\mu(j))  \stackrel{S\to\infty}{\longrightarrow}
\frac{\beta}{2\pi i}\exp(-\beta
l\mu(j)).
\end{equation}
For the  case $l=m=N/2\to\infty$ and
$S=(N-1)/2\to\infty$
the expression for the limiting DF $P_{\infty,l}(\mu(j))$ coincides with (23).
We note that in $S\to\infty$ (or $N\to\infty$) limit, the  continuum
``chemical potentials'' play the role of additional U(1) fluctuating field
whereas for finite $S$ and $N$ they are characterized by fixed and
discrete values.

When $S$ assumes integer values, the minimal fundamental set of
Matsubara frequencies is given by the table in
Fig.\ref{fig:frequencies_spin}.

The exclusion principle for SU(2) in the large spin limit can be also
understood with the help of Fig.\ref{fig:SU(N)_rep} and
Fig.\ref{fig:spin_rep}.  One can see that the empty and the fully occupied states
are canceled in each given circle similarly to even-$N$ SU($N$)
algebra.  The particle-hole (PH) symmetry of the representation results in
an equivalence of single occupied and $2S$ occupied states whereas all
the other states are canceled due to proper weights in the
distribution function (\ref{lsu}). In accordance with PH symmetry
being preserved for each value of the chemical potential all circle diagrams
(see Fig.3, Fig.5) are invariant with respect to simultaneous change
$\mu \leftrightarrow -\mu$ and $n_{particle} \leftrightarrow n_{holes}$.

\subsection{From SU(2) to SO(4)}
We have shown that  the general rectangular Young Tableau of size $n_c*m$ is represented by 
$N*n_c$ component fermionic field with $n_c$ diagonal constraints and $n_c(n_c-1)/2$ 
off-diagonal constraints. However, the fully symmetric representation (one row) requires only
$n_c+1=2S+1$ component field. The general scheme of projected representation for SU(N) group 
is given in \cite{kis01}. We illustrate this idea on a simple example of $S=1$.

We start with $2*n_c=4$ - field representation
\begin{equation}
(a_{11},~ a_{12},~ a_{21},~ a_{22})
\label{1}
\end{equation}
There are two diagonal and two off-diagonal constraints which read as follows:
\begin{equation}
a^\dagger_{11}a_{11}+a^\dagger_{21}a_{21}=1,\;\;\;\;
a^\dagger_{12}a_{12}+a^\dagger_{22}a_{22}=1.
\end{equation}
\begin{equation}
a^\dagger_{11}a_{12}+a^\dagger_{21}a_{22}=0,\;\;\;\;
a^\dagger_{12}a_{11}+a^\dagger_{22}a_{21}=0
\label{so4c1}
\end{equation}
and generators of $SU(2)$ group are given by
$$
S^-=S_2^1=a^\dagger_{11}a_{21}+a^\dagger_{12}a_{22},\;\;\;\;
S^+=S_1^2=a^\dagger_{21}a_{11}+a^\dagger_{22}a_{12}
$$
\begin{equation}
2 S^z=S_2^2-S_1^1=
a^\dagger_{21}a_{21}+a^\dagger_{22}a_{22}-a^\dagger_{11}a_{11}-a^\dagger_{12}a_{12}
\label{sz}
\end{equation}
Combining definition (\ref{sz}) with constraint (\ref{so4c1}) we reach the  following equations:
\begin{equation}
S^-=a^\dagger_{11}(a_{21}+ a_{12})+(a^\dagger_{12}+a^\dagger_{21})a_{22},\;\;\;\;
S^+=(a^\dagger_{21}+ a^\dagger_{12})a_{11}+a^\dagger_{22}(a_{12}+a_{21}),\;\;\;\;\;
S^z=a^\dagger_{22}a_{22}-a^\dagger_{11}a_{11}
\end{equation}
Therefore, we conclude that the antisymmetric (singlet) combination $a_{12}-a_{21}$
does not enter the expression for spin $S=1$ operators. Thus, three (out of four) component
Fermi-field is sufficient for the description of $S=1$ $SU(2)$ representation.
Defining new fields as follows
\begin{equation}
a_{11}=f_{-1},\;\;\;\; a_{22}=f_1,\;\;\;\;\; \frac{1}{\sqrt{2}}(a_{12}+a_{21})=f_0,\;\;\;\;\;\;
\frac{1}{\sqrt{2}}(a_{12}-a_{21})=s.
\label{2}
\end{equation}
where fermions $f_1,f_0,f_{-1}$ stand for $S^z=1,0-1$ projections of the triplet state
and fermion $s$ determines the singlet 
state, we come to standard $S=1$ $SU(2)$ representation (c.f \ref{s1})
\begin{equation}
S^+ =  \sqrt{2}(f_0^\dagger f_{-1}+f^\dagger_{1}f_0),\;\;\;\;
S^- =  \sqrt{2}(f^\dagger_{-1}f_0+ f_0^\dagger f_{1}) ,\;\;\;\;
S_z =  f^\dagger_{1}f_{1}-f^\dagger_{-1}f_{-1},
\label{3}
\end{equation}
with the constraint
\begin{equation}
n_1+n_0+n_{-1}+n_s=2
\label{so4c2}
\end{equation}
where $n_\alpha=f^\dagger_\alpha f_\alpha$. 

Nevertheless, the constraint (\ref{so4c2}) transforms to a standard SU(2) $S=1$ constraint in both cases
$n_s=0$ and $n_s=1$ since there is no singlet/triplet mixing allowed by SU(2) algebra.

To demonstrate the transformation of the local constraint let's first consider the case $n_s=0$.
The constraint reads as follows

\begin{equation}
n_1+n_0+n_{-1}=2S\;\;\;\;\;\iff\;\;\;\;\;\;\;{\bf S}^2=S(S+1).
\end{equation}
On the other hand, the  the states with $2S$ occupation are equivalent to the 
states with single occupation due to particle-hole symmetry. 
Thus, the constraint (\ref{so4c2}) might be written as 
\begin{equation}
\tilde n_1+\tilde n_0+\tilde n_{-1}=1
\end{equation}
where $\tilde n_\alpha =1-n_\alpha$. The latter case corresponds to $n_s=1$. 

We start now with definition of $SO(4)$ group obeying the following commutation relations
\begin{equation}
[S_j,S_k]  = ie_{jkl}S_l,\;\;\;\;\;\;[P_j,P_k]=ie_{jkl}S_l,\;\;\;\;\;\;\;\;\;[P_j,S_k]=ie_{jkl}P_l
\label{coms}
\end{equation}
where 6 generators of SO(4) group, namely vectors ${\bf S}$ and ${\bf P}$ 
are represented by the matrices
\begin{eqnarray}
S^+=\sqrt{2} \left(
\begin{array}{cccc}
  0& 1& 0 & 0\\
  0& 0 & 1& 0 \\
  0& 0& 0& 0 \\
  0& 0& 0& 0
\end{array}
\right),\;\;\;\;
S^-=\sqrt{2} \left(
\begin{array}{cccc}
  0& 0& 0 & 0\\
  1& 0 & 0& 0 \\
  0& 1& 0& 0 \\
  0& 0& 0& 0
\end{array}
\right), \;\;\;\;
S^z=\left(
\begin{array}{cccc}
  1& 0& 0 & 0\\
  0& 0 & 0& 0 \\
  0& 0& -1& 0 \\
  0& 0& 0& 0
\end{array}
\right), \nonumber
\end{eqnarray}
\begin{eqnarray}
P^+=\sqrt{2} \left(
\begin{array}{cccc}
  0& 0& 0 & 1\\
  0& 0 & 0& 0 \\
  0& 0& 0& 0 \\
  0& 0& -1& 0
\end{array}
\right),\;\;\;\;
P^-=\sqrt{2} \left(
\begin{array}{cccc}
  0& 0& 0 & 0\\
  0& 0 & 0& 0 \\
  0& 0& 0& -1 \\
  1& 0& 0& 0
\end{array}
\right),\;\;\;\;
P^z=\left(
\begin{array}{cccc}
  0& 0& 0 & 0\\
  0& 0 & 0& -1 \\
  0& 0& 0& 0 \\
  0& -1& 0& 0
\end{array}
\right). 
\end{eqnarray}
With the Casimir operator
$$
{\bf S\cdot P}=0,\;\;\;\;\;\; {\bf S}^2+ {\bf P}^2=3.
$$
Unlike SU(2) group, the singlet/triplet transitions are allowed in SO(4) group and 
determined by ${\bf P}$ operators.
Using the definition of singlet/triplet fermions one comes to following representation
\begin{equation}
S^+ =  \sqrt{2}(f_0^\dagger f_{-1}+f^\dagger_{1}f_0),\;\;\;\;
S^- =  \sqrt{2}(f^\dagger_{-1}f_0+ f_0^\dagger f_{1}) ,\;\;\;\;
S_z =  f^\dagger_{1}f_{1}-f^\dagger_{-1}f_{-1},
\label{3a}
\end{equation}
\begin{equation}
P^+  =  \sqrt{2}(f^\dagger_{1} s -  s^\dagger f_{-1}), \;\;\;\;
P^-  =  \sqrt{2}(s^\dagger f_{1} - f^\dagger_{-1}s) ,\;\;\;\;
P_z  =  -( f_0^\dagger s + s^\dagger f_0).
\label{4}
\end{equation}
with the only constraint
$$
n_1+n_0+n_{-1}+n_s=1
$$
whereas the orthogonality condition is fulfilled automatically.
\subsection{Real-time formalism}
We discuss finally the real-time formalism based on the semi-fermionic
representation of SU($N$) generators. This approach is necessary for
treating the systems out of equilibrium, especially for many
component systems describing Fermi (Bose) quasiparticles interacting
with spins. The real time formalism \cite{keldysh}, \cite{schwinger} provides an alternative approach
for the analytical continuation method for equilibrium problems
allowing direct calculations of correlators whose analytical
properties as function of many complex arguments can be quite
cumbersome.

\begin{figure}
\begin{center}
  \epsfxsize8cm \epsfbox{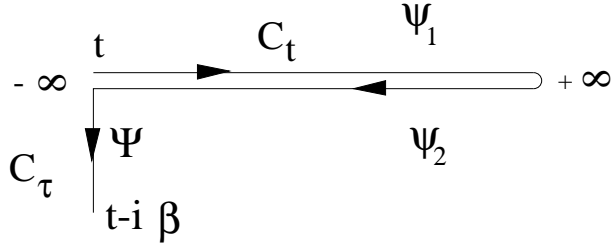} \vspace*{5mm}
\end{center}
\caption{The Keldysh contour going from $-\infty\to\infty\to-\infty$ in real
  time. The boundary conditions on the imaginary time segment
  determine the generalized distribution functions for
  quasiparticles.}
\label{fig:keldysh}
\end{figure}

To derive the real-time formalism for SU($N$) generators we use the
path integral representation along the closed time Keldysh contour
(see Fig.\ref{fig:keldysh}).
Following the standard route \cite{babichenko86a}, we can express the
partition function of the problem containing SU($N$) generators as a
path integral over Grassmann variables
$\psi_l=(a_{l,1}(j),...,a_{l,N}(j))^{T}$ where $l=1,2$ stands for
upper and lower parts of the Keldysh contour, respectively,
\begin{equation}
{\cal Z}/{\cal Z}_0=\int D\bar\psi D\psi\exp(i{\cal A})/
\int D\bar\psi D\psi\exp(i{\cal A}_0)
\label{pfunk}
\end{equation}
where the actions ${\cal A}$ and ${\cal A}_0$ are taken as an integral
along the closed-time contour $C_t+C_\tau$ which is shown in Fig.\ref{fig:keldysh}.
The contour is closed at $t=-\infty+i\tau$ since $\exp(-\beta
H_0)=T_\tau\exp\left(-\int_0^\beta H_0 d\tau\right).$ We denote the
$\psi$ fields on upper and lower sides of the contour $C_t$ as
$\psi_1$ and $\psi_2$ respectively. The fields $\Psi$ stand for the
contour $C_\tau$. These fields provide the matching conditions for
$\psi_{1,2}$ and are excluded from the final expressions.  Taking into
account the semi-fermionic boundary conditions for generalized
Grassmann fields (\ref{bk}) one gets the matching conditions for
$\psi_{1,2}$ at $t=\pm\infty$,
$$
\psi^\mu_{1,\alpha}|_k(-\infty) =
  \exp\left(i\pi\frac{2k-1}{N}\right)\psi^\mu_{2,\alpha}|_k(-\infty),
$$
\begin{equation}
  \psi^\mu_{1,\alpha}|_k(+\infty) =\psi^\mu_{2,\alpha}|_k(+\infty)
\end{equation}
for $k=1,...,\lfloor N/2 \rfloor$ and $\alpha=1,...,N$.  The
correlation functions can be represented as functional derivatives of
the generating functional
\begin{equation}
Z[\eta]={\cal Z}_0^{-1}\int D\bar\psi D\psi\exp\left(i{\cal A}+
i\oint_C d t (\bar\eta\sigma^z\psi+\bar\psi\sigma^z \eta)\right)
\end{equation}
where $\eta$ represents sources and the $\sigma^z$ matrix stands for
"causal" and "anti-causal" orderings along the contour.

The on-site Green's functions (GF) which are matrices of size
$2N\times 2N$ with respect to both Keldysh (lower) and spin-color
(upper) indices are given by
\begin{equation}
G_{\mu\nu}^{\alpha\beta}(t,t')=
-i\frac{\delta}{i\delta\bar \eta_\mu^\alpha(t)}
\frac{\delta}{i\delta \eta_\nu^\beta(t')}
Z[\eta]|_{\bar\eta,\eta\to 0}.
\label{rtf}
\end{equation}
To distinguish between imaginary-time (\ref{itf}) and
real-time (\ref{rtf}) GF's,
we use different notations for Green's functions in these representations.

After a standard shift-transformation \cite{babichenko86a} of the fields $\psi$
the Keldysh GF of free semi-fermions assumes the form
\begin{eqnarray}
G_0^\alpha(\epsilon)=G^{R,\alpha}_0
\left(
\begin{array}{cc}
1 - f_\epsilon &  -f_\epsilon\\
1 - f_\epsilon &  -f_\epsilon
\end{array}\right)-
G^{A,\alpha}_0
\left(
\begin{array}{cc}
-f_\epsilon & -f_\epsilon\\
1 - f_\epsilon & 1 - f_\epsilon
\end{array}
\right),
\nonumber
\end{eqnarray}
where the retarded and advanced GF's are
\begin{equation}
G^{(R,A)\alpha}_0(\epsilon)=(\epsilon \pm i\delta)^{-1},
\quad
f_\epsilon=f^{(N,k)}(\epsilon),
\end{equation}
with equilibrium distribution functions
\begin{equation}
f^{(N,k)}(\epsilon)=T\sum_n\frac{e^{i\omega_{n_k}\tau|_{+0}}}
{i\omega_{n_k}-\epsilon}=
\frac{1}{e^{i\pi (2k-1)/N}\exp(\beta\epsilon)+ 1}.
\end{equation}
A straightforward calculation of $f^{(N,k)}$ for the case of even $N$
leads to the following expression
\begin{equation}
f^{(N,k)}(\epsilon)
=\frac{\displaystyle\sum_{l=1}^N(-1)^{l-1}
\exp\left(\beta\epsilon(N-l)\right)
\exp\left(-\frac{i\pi l(2k-1)}{N}\right)}{\exp(N\beta\epsilon)+1},
\end{equation}
where $k=1,...,N/2.$
The equilibrium distribution functions (EDF) $f^{(2S+1,k)}$ for the
auxiliary Fermi-fields representing arbitrary $S$ for $SU(2)$ algebra
are given by
\begin{equation}
f^{(2S+1,k)}(\epsilon)
=\frac{\displaystyle\sum_{l=1}^{2S+1}(-1)^{l-1}
\exp\left(\beta\epsilon(2S+1-l)\right)
\exp\left(-\frac{i\pi(2k-1)}{2S+1})\right)}
{\exp((2S+1)\beta\epsilon)+(-1)^{2S+1}}
\end{equation}
for $k=1,...,\lfloor S+1/2 \rfloor$.  Particularly simple are the
cases of $S=1/2$ and $S=1$,
$$
 f^{(2,1)}(\epsilon)=
    n_F(2\epsilon)-i\frac{1}{2\cosh(\beta\epsilon)}
$$
\begin{equation}
     f^{(3,1)}(\epsilon)=
    \frac{1}{2}n_B(\epsilon)-\frac{3}{2}n_B(3\epsilon)
    -i\sqrt{3}\frac{\sinh(\beta\epsilon/2)}{\sinh(3\beta\epsilon/2)}
\end{equation}
Here, the standard notations for Fermi/Bose distribution functions
$n_{F/B}(\epsilon)=[\exp(\beta\epsilon) \pm 1]^{-1}$ are used. For $S=1/2$ the 
semi-fermionic EDF satisfies the
obvious identity $|f^{(2,1)}(\epsilon)|^2=n_F(2\epsilon)$.

In general the EDF for half-integer and integer spins can be expressed in
terms of Fermi and Bose EDF respectively.  We note that since
auxiliary Fermi fields introduced for the representation of SU($N$)
generators do not represent the true quasiparticles of the problem,
helping only to treat properly the constraint condition, the
distribution functions for these objects in general do not have to be
real functions.  Nevertheless, one can prove that the imaginary part
of the EDF does not affect the physical correlators and can be
eliminated by introducing an infinitesimally small real part for the
chemical potential.  In spin problems, a uniform/staggered magnetic
field usually plays the role of such real chemical potential for
semi-fermions.

\section{Application of semi-fermionic representation}
In this section we illustrate some of the applications of SF representation for various
problems of strongly correlated physics.

\subsection{Heisenberg model: FM, AFM and RVB}

The effective nonpolynomial action for Heisenberg model with ferromagnetic (FM) coupling
has been investigated in \cite{popov88a}. The model with antiferromagnetic (AFM) interaction has been
considered by means of semi-fermionic representation in \cite{kiselev99a} and \cite{azakov} 
(magnon spectra) and in \cite{kiselev00b} for
resonance valence bond (RVB) excitations. The Hamiltonian considered is given as
\begin{equation}
H_{int}=-\sum_{<ij>}J_{ij}\left(\vec{S}_i\vec{S}_j-\frac{1}{4}\right)
\label{hh1}
\end{equation}
\begin{itemize}
\item Ferromagnetic coupling $J=I_{FM}>0$
\end{itemize}
The exchange  $\vec{S}_i\vec{S}_j$ is represented as 
four-semi-fermion interaction. Applying the Hubbard-Stratonovich
transformation by the {\it local vector} field $\vec{\Phi}_i(\tau)$ the effective nonpolynomial 
action is obtained in terms of vector c-field. The FM phase transition corresponds to the
appearance at $T\leq T_c$ of the nonzero average $\langle \Phi^z(q=0,0)\rangle$ which stands for
the nonzero magnetization, or in other words, corresponds to the Bose condensation
of the field $\Phi^z$. 
\begin{equation}
\Phi^z(\vec{k},\omega)={\cal
M}(\beta N)^{1/2}\delta_{\vec{k},0} \delta_{\omega,0}
+\tilde\Phi^z(\vec{k},\omega).
\end{equation}
In one loop approximation the
standard molecular field equation can be reproduced
\begin{equation}
{\cal M}=I_{FM}(0)\tanh(\beta{\cal M}/2).
\label{n1}
\end{equation}
The saddle point (mean-field) effective action is given by
well-known expression
\begin{equation}
{\cal A}_0[{\cal M}]=-N\left[\frac{\beta {\cal M}^2}{4I_M(0)}-
\ln\left(2\cosh\left(\frac{\beta {\cal M}}{2}\right)\right)\right],
\end{equation}
and the free energy per spin $f_0$  is determined by the standard
equation:
\begin{equation}
\beta f_0 =-\ln Z_S=\frac{\beta {\cal M}^2}{4I_M(0)}-
\ln\left(2\cosh\left(\frac{\beta {\cal M}}{2}\right)\right)
\end{equation}
Calculation of the second variation of ${\cal A}_{eff}$ gives rise to
the following expression
$$\delta{\cal A}_{eff}=-\frac{1}{4}\sum_{\vec{k}}\Phi^z(\vec{k},0)
\left[I_M^{-1}(\vec{k})-\frac{\beta}{2\cosh^2(\beta\Omega)}\right]
\Phi^z(\vec{k},0)
-\frac{1}{4}\sum_{\vec{k},\omega\ne 0}
I_M^{-1}(\vec{k})\Phi^z(\vec{k},\omega)\Phi^z(\vec{k},\omega)-$$
\begin{equation}
-\sum_{\vec{k},\omega}\Phi^+(\vec{k},\omega)\left[
I_M^{-1}(\vec{k})-\frac{\tanh(\beta\Omega)}{2\Omega - i\omega}\right]
\Phi^-(\vec{k},\omega)
\end{equation}
where $\Omega=(g\mu_B H +{\cal M})/2$. The magnon spectrum ($T\leq T_c$) is 
determined by the poles of $\langle \Phi^+\Phi^-\rangle$ correlator, $\omega=\lambda {\bf k}^2$.
\begin{itemize}
\item Antiferromagnetic coupling $J=I_{AFM}<0$. {\it N\'eel solution}
\end{itemize}
The AFM transition corresponds to formation of the staggered condensate
\begin{equation}
\Phi^z(\vec{k},\omega)={\cal N}(\beta N)^{1/2}\delta_{\vec{k},\vec{Q}} \delta_{\omega,0}
+\tilde\Phi^z(\vec{k},\omega)
\end{equation}
The one-loop approximation leads to standard mean-field equations for the staggered magnetization
\begin{equation}
{\cal N}=-I_{AFM}(Q)\tanh(\beta{\cal N}/2),
\;\;\;\;\;\;
{\cal A}_0[{\cal N}]=N\left[\frac{\beta {\cal N}^2}{4I_{AFM}(Q)}+
\ln\left(2 \cosh\left( \frac{\beta {\cal N}}{2}\right)\right)\right].
\end{equation}

After taking into account the second variation of ${\cal A}_{eff}$,
the following expression for the effective action is obtained [(see e.g.
\cite{kiselev99a},\cite{azakov}):
$$\delta{\cal A}_{eff}=\frac{1}{4}\sum_{\vec{k}}\Phi^z(\vec{k},0)
\left[I_{AFM}^{-1}(\vec{k})+\frac{\beta}{2\cosh^2(\beta\tilde\Omega)}\right]
\Phi^z(\vec{k},0)
+\frac{1}{4}\sum_{\vec{k},\omega\ne 0}
I_{AFM}^{-1}(\vec{k})\Phi^z(\vec{k},\omega)\Phi^z(\vec{k},\omega)+$$
\begin{equation}
+\sum_{\vec{k},\omega}\Phi^+(\vec{k},\omega)\left[
I_{AFM}^{-1}(\vec{k})+\frac{2\tilde\Omega\tanh(\beta\tilde\Omega)}
{4\tilde\Omega^2 +\omega^2}
\right]\Phi^-(\vec{k},\omega)
-\sum_{\vec{k},\omega}\Phi^+(\vec{k}+\vec{Q},\omega)
\frac{i\omega}{4\tilde\Omega^2 +\omega^2}
\Phi^-(\vec{k},\omega).
\end{equation}
The AFM magnon spectrum $\omega= c|{\bf k}|$.
\begin{itemize}
\item Antiferromagnetic coupling. {\it Resonance Valence Bond solution}
\end{itemize}
The four-semi-fermion term in (\ref{hh1}) is decoupled by {\it bilocal scalar} field $\Lambda_{ij}$.
The RVB spin liquid (SL) instability in 2D Heisenberg model corresponds to 
Bose-condensation of exciton-like \cite{com1} pairs of semi-fermions:
\begin{equation}
\Delta_0=-\sum_{\bf q}\frac{I_{\bf q}}{I_0}\tanh\left(\frac{I_{\bf q}\Delta_0}{T}\right),
\;\;\;\;\;
{\cal A}_0=\frac{\beta |I|\Delta_0^2}{2}-\sum_{\bf q}\ln\left[2 \cosh(\beta I_{\bf q}\Delta_0)\right]
\label{rvb}
\end{equation}
where $\Delta_0=\Delta({\bf q}=0)$ is determined by the modulus of $\Lambda_{ij}$ field
\begin{equation}
\Lambda_{<ij>}(\vec{R},\;\vec{r})=
\Delta(\vec{r})\exp\left(i\vec{r}\vec{A}(\vec{R})\right)
\label{uni}
\end{equation}
whereas the second variation of $\delta {\cal A}_{eff}$ describes  the fluctuations of phase $\Lambda_{ij}$
\begin{equation}
{\cal A}_{eff}=\sum_{{\bf k},\omega}
A_\alpha({\bf k},\omega)\pi^{\alpha\beta}_{{\bf k},\omega}A_\beta({\bf k},\omega),\;\;\;\;\;\;
{\Large \pi}^{\alpha\beta}_{{\bf k},\omega}=Tr(p^\alpha p^\beta
(G_{p+k}G_p+G_{p+k}G_p)+\delta_{\alpha\beta}f(I_{\bf p}\Delta_0))
\end{equation}
The spectrum of excitation in uniform SL is determined by zeros of $\pi^R$ and is purely diffusive
\cite{ioffe}-\cite{lee}. 

\subsection{Kondo lattices: competition between magnetic and Kondo correlations}
The problem of competition between Ruderman-Kittel-Kasuya-Yosida (RKKY)
magnetic exchange and Kondo correlations is one of the most interesting
problem of the heavy fermion physics. The recent experiments unambiguously show,
that such a competition  is responsible for many unusual properties of
the integer valent heavy fermion compounds e.g. quantum critical 
behavior, unusual antiferromagnetism  and superconductivity (see references in \cite{kis02a}).
We address the reader to the review \cite{col} for details of  complex physics of Kondo effect
in heavy fermion compounds. In this section we discuss the influence of Kondo effect on the 
competition between local (magnetic, spin glass) and non-local (RVB) correlations.
The Ginzburg-Landau theory for nearly antiferromagnetic Kondo lattices has been constructed in 
\cite{kis02a}
using the semi-fermion approach. We discuss the key results of this theory.

The  Hamiltonian of the Kondo lattice (KL) model is given by 
\begin{equation}
H=\sum_{k\sigma}\varepsilon_k c^\dagger_{k\sigma}c_{k\sigma}+
J\sum_{j}\left({\bf S}_j{\bf s}_j+\frac{1}{4}N_jn_j
\right)
\label{2.1}
\end{equation}
Here the local electron and spin density operators for conduction electrons 
 at site $j$ are defined as 
\begin{equation}
n_j=\sum_{j\sigma}c^\dagger_{j\sigma}c_{j\sigma},~~~ 
{\bf s}_j=\sum_{\sigma}\frac{1}{2}c^\dagger_{j\sigma}
{\hat\tau}_{\sigma\sigma'}c_{j\sigma'}, 
\label{2.2}
\end{equation}
where ${\hat\tau}$ are the Pauli 
matrices and $c_{j\sigma}=\sum_k c_{k\sigma}\exp (ikj)$.
The spin glass (SG) freezing is possible if an additional quenched randomness
of the inter-site exchange $I_{jl}$ between the localized spins arises. 
This disorder is described by 
\begin{equation}
H'=\sum_{jl}I_{jl}({\bf S}_j{\bf S}_l).
\label{2.3}
\end{equation}
We start with a perfect Kondo lattice. The spin correlations in KL are
characterized by two energy scales, i.e., $I\sim \ J^2/\varepsilon_F,$ and 
$\Delta_K\sim \varepsilon_F\exp(-\varepsilon_F/J)$ 
(the inter-site indirect exchange of the RKKY type
and the Kondo binding energy, respectively). At high enough temperature, the
localized spins are weakly coupled with the electron Fermi sea 
having the Fermi energy $\varepsilon_F$, so that the magnetic
response of a rare-earth sublattice of KL is of paramagnetic Curie-Weiss type.
With decreasing temperature either a crossover to a strong-coupling 
Kondo singlet regime occurs at $T \sim \Delta_K$ or the phase transition 
to an AFM state occurs at $T=T_N \sim zI$ where $z$ is a coordination number
in KL. If $T_N \approx \Delta_K$ the interference between two trends results
in the decrease of both characteristic temperatures 
or in suppressing one of them.
The mechanism of suppression is based on the screening effect due to Kondo interaction.
As we will show, the Kondo correlations screen the local order parameter, but leave nonlocal
correlations intact. The mechanism of Kondo screening for single-impurity Kondo problem is
illustrated on Fig.5
\vspace*{-15mm}
\begin{figure}
\begin{center}
  \epsfysize6cm \epsfbox{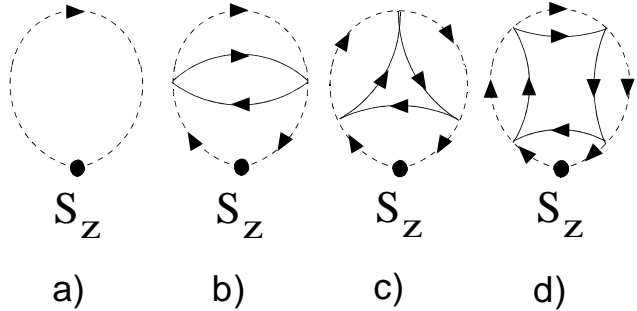} 
\vspace*{-10mm}
\end{center}
\vspace*{2mm}
\caption{Kondo screening of the local moment in single-impurity Kondo problem. Dashed line denotes semi-fermions, solid line stands for conduction electrons.}
\label{fig:pol2}
\end{figure}
As a result, the magnetization of local impurity in the presence of Kondo effect is 
determined in terms of GF's of semi-fermions ${\cal G}(\omega)$  by the following expression \cite{WT}:
\begin{equation}
{\cal M}(H)=S(g\mu_B)T\sum_{\omega}\left({\cal G}_\uparrow(\omega)-
{\cal G}_\downarrow(\omega)\right)=S(g\mu_B)
\tanh\left(\frac{H\beta}{2}\right)
\left[1-\frac{1}{\ln(T/T_K)}-\frac{\ln(\ln(T/T_K))}{2\ln^2(T/T_K)}+...\right].
\end{equation}
To take into account the screening effect in the lattice model we apply the semi-fermionic representation
of spin operators. In accordance with the general path-integral approach to KL's, we first integrate 
over fast (electron) degrees of freedom. The Kondo exchange interaction is decoupled by
auxiliary field $\phi$ \cite{read83a} 
with statistics complementary to that of semi-fermions which prevents this field
from Bose condensation except at $T=0$. As a result, we are left with an effective bosonic action
describing low-energy properties of KL model at high $T>T_K$ temperatures.
\begin{itemize}
\item Kondo screening of the N\'eel order
\end{itemize}
To analyze the influence of Kondo screening on formation of AFM order, we adopt the decoupling 
scheme for the Heisenberg model discussed in Section II.A. Taking into account the classic part
of N\'eel field, we calculate the Kondo-contribution to the effective action which depends on magnetic
order parameter ${\cal N}$:
\begin{equation}
{\cal A}_{\phi}=2\sum_{{\bf q},n}
\left[\frac{1}{\widetilde J}-\Pi({\cal N})\right] 
|\phi_n({\bf q})|^2.
\label{3.1}
\end{equation}
where a polarization operator $\Pi({\cal N})$ casts the form
\begin{equation}
\Pi({\cal N})=\rho(0)\ln\left(\frac{\epsilon_F}{T}\right)+\left[
\frac{\pi}{2}\left(\frac{1}{\cosh(\beta{\cal N})}-1\right)
+O\left(\frac{{\cal N}^2}{T\epsilon_F}\right)\right]~,
\label{3.2}
\end{equation}
where $\rho(0)$ is the density of states of conduction electrons at the Fermi level and the
Kondo temperature $T_K=\epsilon_F\exp\left(-1/(\rho(0)J)\right)$.
Minimizing the effective action ${\cal A}(\phi, {\cal N})$ with respect 
to classic field ${\cal N}$, the mean field equation
for N\'eel transition is obtained (c.f. with (\ref{n1}))
\begin{equation}
{\cal N}=\tanh\left(\displaystyle\frac{I_{\bf Q}{\cal N}}{2T}\right)
\left[\displaystyle
1-\frac{a_N}{\displaystyle\ln\left(T/T_K\right)}
\frac{\cosh^2(\beta I_{\bf Q}{\cal N}/2)} 
{\cosh^2(\beta I_{\bf Q} {\cal N})}\right].
\label{3.4}
\end{equation}
As a result, Kondo corrections to the molecular field equation reduce the N\'eel temperature
\begin{itemize}
\item Kondo enhancement of RVB correlations
\end{itemize}
Applying the similar procedure to nonlocal RVB correlations, we take into account the influence
of Kondo effect on RVB correlations
\begin{equation}
\Pi(I_{\bf q}\Delta)=\rho(0)\ln\left(\frac{\epsilon_F}{T}\right)+\sum_{\bf k}
\left[\frac{1}{\cosh \beta(I_{\bf k}\Delta)}-1+ 
I_{\bf k}\Delta\tanh(\beta I_{\bf k}\Delta)\right]
\frac{1}{\xi^2_{\bf k+q}+ (\pi/2\beta)^2}~.
\label{3.6}
\end{equation}
Here $\xi_k=\epsilon({\bf k})-\epsilon_F$.
Minimizing the effective action with respect to $\Delta$ we obtain new self-consistent equation
to determine the non-local semi-fermion correlator.
\begin{equation}
\Delta= -\sum_{\bf q}\frac{I_{\bf q}}{I_0} \left[\tanh\left(\frac{I_{\bf q}\Delta}{T}
\right)+a_{sl}\frac{I_{\bf q} \Delta}{T\ln(T/T_K)}\right].
\label{3.4a}
\end{equation}
It is seen that unlike the case of local magnetic order, the Kondo scattering favors transition into
the spin-liquid state, because the scattering means the  involvement of the itinerant 
electron degrees of freedom into the spinon dynamics.
\begin{itemize}
\item Kondo effect and quenched disorder
\end{itemize}
Let's assume that the RKKY interactions are random (e.g. due to the presence of non-magnetic impurities
resulting in appearance of random phase in the 
RKKY indirect exchange). In this case the spin glass phase
should be considered. As it has been shown in \cite{kiselev00a} and \cite{kis02a}, the 
influence of static disorder on Kondo effect in models with Ising exchange 
on fully connected lattices (Sherrington-Kirkpatrick model) can be taken into account by the mapping 
KL model with quenched disorder
onto the single impurity Kondo model in random (depending on replicas) 
magnetic field. It allows for the
self-consistent determination of the Edwards-Anderson $q_{EA}$order parameter given by the 
following set
of self-consistent equations
\begin{equation}
\tilde q  =  1-\frac{2c}{\ln(T/T_K)} -
O\left(\frac{1}{\ln^2(T/T_K)}\right),\;\;\;\;
q  =  \int_x^G\tanh^2\left(\frac{\beta I x \sqrt{q}} {1+2c(\beta
I)^2(\tilde q - q)/\ln(T/T_K)}
\right)+O\left(\frac{q}{\ln^2(T/T_K)}\right) ~. 
\end{equation}
Here $q=q_{EA}$ and $\tilde q$ are nondiagonal and diagonal elements of Parisi matrix respectively.
Therefore, the Kondo-scattering  results in the depression of the freezing temperature due to the 
screening effects in the same way as the magnetic moments and the 
one-site susceptibility are screened in the
single-impurity Kondo problem (c.f. Fig.5)  when Ising and Kondo interactions are of the same order of magnitude.
\begin{figure}
\begin{center}
\epsfxsize35mm
\epsfbox{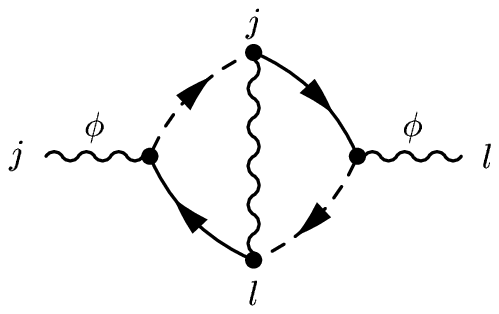}
\epsfxsize35.mm
\epsfbox{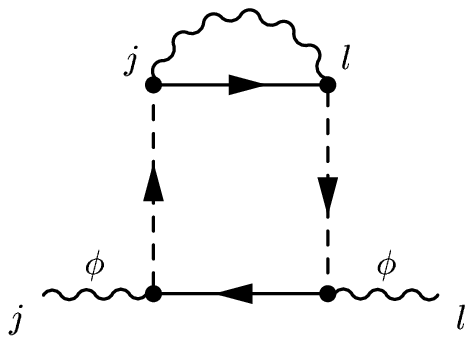}
\caption{Feynman diagrams for nonlocal excitations associated with the overlap
of Kondo clouds.}
\end{center}
\end{figure}
Let's now briefly discuss the  fluctuation effects in Kondo lattices. The natural way to construct 
the fluctuation theory is to consider the non-local dynamical Kondo correlations described by the field 
$\phi({\bf q},\omega)$ (see Fig.6). In fact, the non-locality of the ``semi-Bosonic'' field is associated with
an overlap of Kondo clouds \cite{kis02a} and responsible for a crossover from the localized magnetism to
the itinerant-like fluctuational spin-liquid magnetism. The temperature dependence of static magnetic 
susceptibility  becomes nonuniversal in spite of the fact that we are in a region of critical 
AFM fluctuations which is consistent with recent experimental observations.

\subsection{Kondo effect in quantum dots}
The single electron tunneling through the 
quantum dot \cite{Glaz} has been  studied in great details during the recent decade. 
Among many interesting phenomena
behind the unusual transport properties of mesoscopic systems,
the Kondo effect in quantum dots, recently observed experimentally,
continues to attract an attention both of experimental and 
theoretical communities.  The modern nanoscience technologies 
allow one to produce the highly controllable systems based on 
quantum dot devices and possessing many of properties of strongly 
correlated electron systems. The quantum dot in a semiconductor planar 
heterostructure is a confined few-electron system (see Fig.7) contacted by sheets of 
two-dimensional gas (leads). Junctions between dot and leeds produce the
exchange interaction between the spins of the dot and spins of itinerant 2D
electron gas. Measuring the dc $I-V$ characteristics, one can investigate  
the Kondo effect in quantum dots under various conditions.

Various realizations of Kondo effect in quantum dots were proposed both theoretically
and experimentally in recent publications (see e.g. \cite{KA02}  for review). In order to illustrate the 
application of semi-fermionic approach we discuss briefly electric field induced Kondo
tunneling in double quantum dot (DQD).
As was noticed in \cite{KA01}, quantum dots with even ${\cal N}$ possess
the dynamical symmetry $SO(4)$ of spin rotator in the Kondo tunneling regime,
provided the low-energy part of its spectrum is formed by a singlet-triplet
(ST) pair, and all other excitations are separated from the ST manifold by
a gap noticeably exceeding the tunneling rate $\gamma$. A DQD with 
even ${\cal N}$ in a side-bound configuration
where two wells are coupled by the tunneling $v$ and only 
one of them (say, $l$) is coupled to metallic 
leads $(L,R)$ is a simplest system satisfying this condition
\cite{KA01}. Such system was realized experimentally in Ref.\cite{mol95}.
\begin{figure}
\begin{center}
\epsfxsize42mm
\epsfysize28mm
\epsfbox{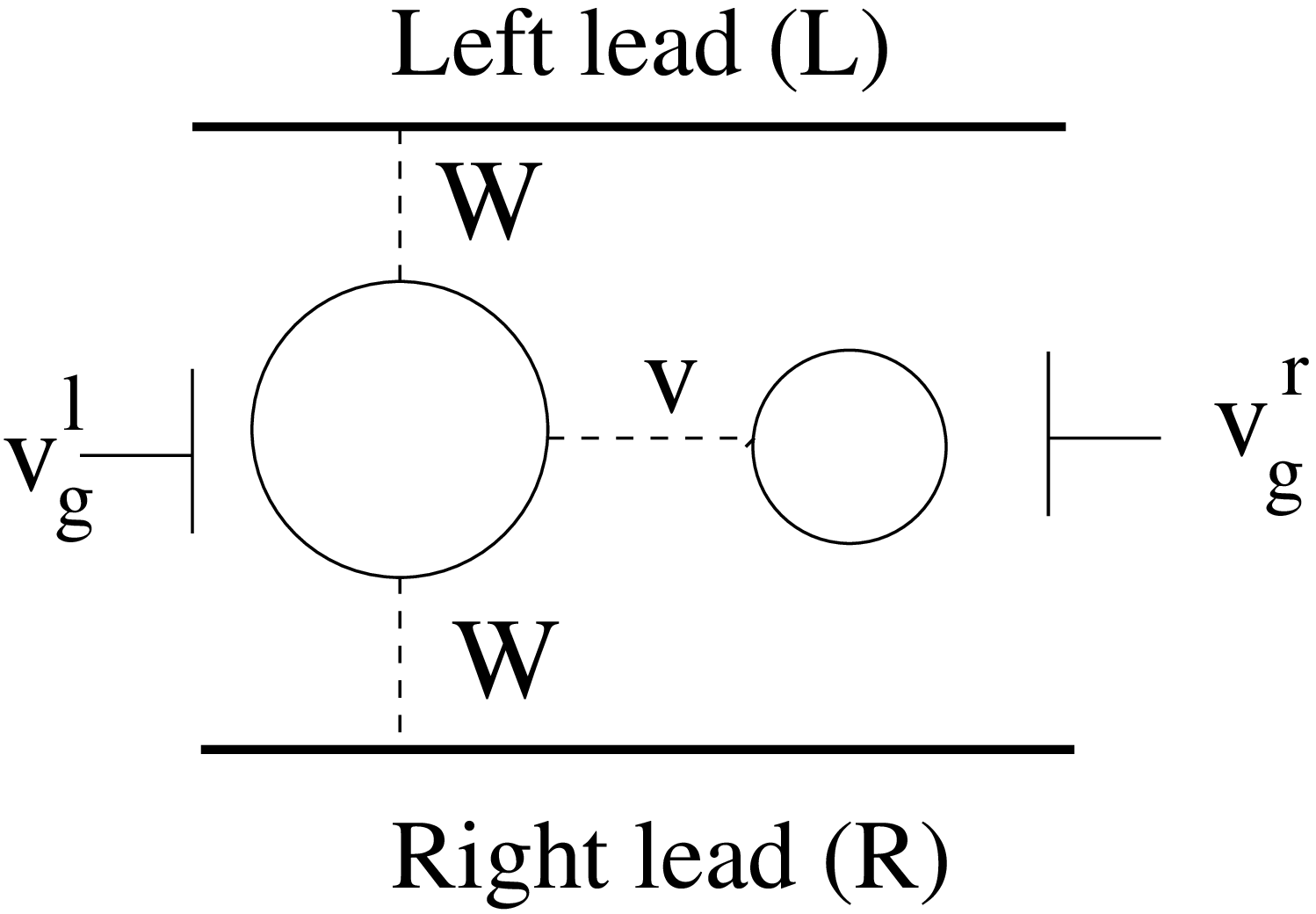}
\hspace*{15mm}
\epsfxsize42mm
\epsfbox{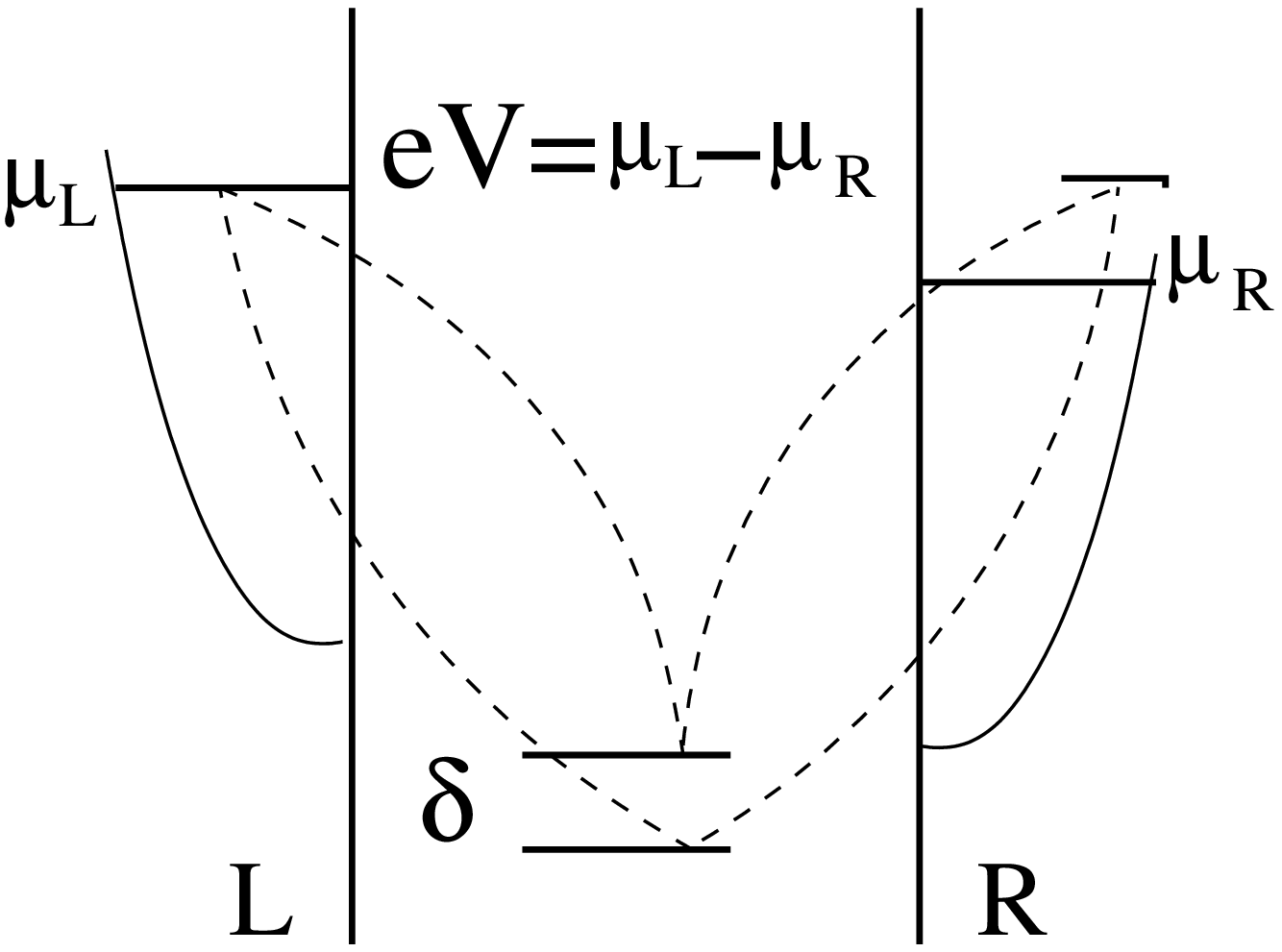}
\mbox{}\\
a)\hspace*{50mm} b)\\
\caption{(a) Double quantum dot in a side-bound configuration
(b) co-tunneling processes 
in biased DQD responsible for the resonance Kondo tunneling.}
\end{center}
\end{figure}
As it was shown in \cite{kis02b} the Shrieffer-Wolff  (SW) transformation, when applied
to a spin rotator results in the following effective spin Hamiltonian
\begin{equation}
H_{int}=\sum_{kk',\alpha\alpha'=L,R}J^S_{\alpha\alpha'}
f_s^\dagger f_s c^\dagger_{k\alpha\sigma}c_{k'\alpha'\sigma}
+\sum_{kk',\alpha\alpha'\Lambda\Lambda'}\left(J^T_{\alpha\alpha'}
\hat S^d_{\Lambda \Lambda'}
+J^{ST}_{\alpha\alpha'}\hat P^d_{\Lambda \Lambda'}\right)\tau^d_{\sigma\sigma'}
c^\dagger_{k\alpha\sigma}c_{k'\alpha'\sigma'}f_\Lambda^\dagger f_{\Lambda'}
\label{hint}
\end{equation}
where the $c$-operators describe the electrons in the leads and $f$-operators 
stand for the electrons in the dot.
The matrices  $\hat S^d$ and $\hat P^d$ ($d$$=$$x$,$y$,$z$) are $4\times 4$ 
matrices  defined by relations (\ref{coms}) (see Section I.C)
and $J^S=J^{SS}$, $J^T=J^{TT}$ and $J^{ST}$ are singlet, triplet and singlet-triplet 
coupling SW constants, respectively.

Applying the semi-fermionic representation of SO(4) group introduced in Section I.C  we 
started with perturbation theory results analyzing the most divergent Feynman diagrams
for spin-rotator model \cite{kis02b}. Following  the ``poor man's scaling'' approach we derive
the system of  coupled renormalization group equations for effective couplings responsible
for the transport through DQD.
As a result, the differential conductance $G(eV,T)/G_0\sim |J^{ST}_{LR}|^2$ 
is shown to be the universal function of 
two parameters $T/T_K$ and $V/T_K$, $G_0=e^2/\pi \hbar$:
\begin{equation}
G/G_0 \sim \ln^{-2}\left(\max[(eV-\delta),T]/T_K\right)
\label{dcond}
\end{equation}
Thus,  the tunneling through singlet DQDs with 
$\delta=E_T-E_S\gg T_K$ exhibits a peak in 
differential conductance at $eV\approx \delta$ 
instead of the usual zero bias Kondo anomaly which arises in 
the opposite limit, $\delta < T_K$.
Therefore, in this case the Kondo effect in DQD is induced by 
a strong external bias. The scaling equations can also be derived in 
Schwinger-Keldysh formalism (see \cite{kiselev00b} and also \cite{kis01}) by
applying the ``poor man's scaling'' approach directly to the dot conductance. 
The detailed analysis of the model (\ref{hint}) 
in a real-time formalism is a subject for a separate publication.

\section{Epilogue and perspectives}
In this paper, we demonstrated several examples of the
applications of semi-fermionic representation to various problems
of condensed matter physics. The list of these applications is not exhaustive. We did not discuss, e.g.,
the interesting development of SF approach for the Hubbard model with repulsive \cite{gros90a} 
and attractive \cite{oppermann91a}
interaction, Dicke model,  2D Ising model in transverse magnetic field, 
application of SF formalism to mesoscopic physics \cite{col02} etc.
Nevertheless, we would like to point out  some problems of strongly correlated physics
where the application of SF representation might be a promising alternative to existing field-theoretical 
methods.\\
\mbox{}\\
{\it Heavy Fermions}
\begin{itemize}
\item Crossover from localized to itinerant magnetism in Kondo lattices
\item Quantum critical phenomena associated with competition between local and nonlocal correlations
\item Nonequilibrium spin liquids
\item Effects of spin impurities and defects in spin liquids
\item Crystalline Electric Field excitations in spin liquids
\item Dynamic theory of screening effects in Kondo spin glasses.
\end{itemize}
{\it Mesoscopic systems}
\begin{itemize}
\item Nonequilibrium Kondo effect in Quantum Dots
\item Two-channel Kondo in complex multiple dots
\item Spin chains, rings and ladders
\item Nonequilibrium spin transport in wires
\end{itemize}
Summarizing, we constructed a general concept of semi-fermionic representation for SU(N) groups.
The main advantage of this representation in application to the strongly correlated systems in 
comparison with another methods is that the local constraint is taken into account exactly and the usual
Feynman diagrammatic codex is applicable. The method proposed allows us to treat spins on the same
footing as Fermi and Bose systems. The semi-fermionic approach can be helpful for the description of
the quantum systems in the vicinity of a quantum phase transition point and for the 
nonequilibrium spin systems. 
\section*{Acknowledgments}
I am grateful to my colleagues F.Bouis, H.Feldmann, K.Kikoin and R.Oppermann
for fruitful collaboration on different stages of SF project. I am thankful to Alexander
von Humboldt Foundation for financial support. The support of Deutsche Forschungsgemeinschaft 
(SFB-410 project) is gratefully acknowledged. 
My special thank to participants of Strongly Correlated Workshops
in Trieste and especially to A.Protogenov for many inspiring discussions.
My particular thank to A.Dutta for careful reading of this manuscript and useful suggestions.

\end{document}